\documentclass[12pt,a4paper]{article}
\usepackage[dvips]{feynmp}
\usepackage{latexsym}
\def\slash#1{#1\hskip -0.5em/}

\newcommand{\tr}[1]{{\rm tr}\left[ #1 \right]}
\newcommand{\ms}{$\overline{\rm MS}$}

\begin{document}

\thispagestyle{empty}

\begin{flushright}
DESY 95-075\\
hep-ph/9505309\\[.5\baselineskip]
May 1995
\end{flushright}

\vspace*{1cm}
\begin{center}
{\Large\bf Renormalized Soft-Higgs Theorems}\\
\vspace*{1cm}
{\sc Wolfgang Kilian}\\
\vspace*{5mm}
Theory Group, DESY \\
22603 Hamburg, Germany
\end{center}

\vfill
\centerline{\bf Abstract}
\vspace*{\baselineskip}
\noindent
The Higgs couplings to matter fields are proportional to their masses.
Thus Higgs amplitudes can be obtained by differentiating amplitudes
without Higgs with respect to masses.  We show how this well-known
statement can be extended to higher order when renormalization effects
are taken into account.  We establish the connection with the
Callan-Symanzik and renormalization group equations and consider also
pseudoscalar Higgs couplings to fermions.  Furthermore, we address the
case where the Higgs couples to a heavy particle that is integrated
out from the low-energy effective Lagrangian.  We derive effective
interactions where mass logarithms are resummed by
renormalization-group methods, and give expansions of the results up
to next-to-leading order.
\newpage
\setcounter{page}{1}

\section{Introduction}

The Higgs boson provides a simple mechanism to accommodate massive
vector bosons and fermions in the standard model.  Present-day
experiments, together with calculations of Higgs interactions up to
two-loop order in some cases, have been used to establish lower
limits for Higgs masses approaching the $W$ and $Z$ mass range, and
the next generation of colliders may give a definite answer to the
question of its existence.

A widely used tool in the study of Higgs interactions are low-energy
theorems (soft-Higgs theorems), which play a role comparable to the
low-energy theorems for pion amplitudes in hadronic physics.  They
rely on the fact that the explicit breaking of scale invariance by the
Higgs interactions can be employed to relate tree-level amplitudes
with different numbers of zero-momentum Higgs fields~\cite{EGN76}.
This theorem has been extended to one-loop amplitudes, where scale
invariance becomes anomalously broken, and it has been observed that
there exists some connection with the scaling functions (beta
functions) of renormalization group theory~\cite{VVSS79}.  Various
applications can be found in the
literature~\cite{SHT-applications,HHG}, and recently with its help
two-loop amplitudes were calculated in the heavy-top
limit~\cite{D91,KS94a,KS94b,KS95}, where algorithms were devised to
take into account the renormalization effects.  However, the role of
scale anomalies in higher order has remained unclear in the present
context~\cite{Sh89}, and thus the precise form of the theorem for
renormalized amplitudes in the general case has remained unknown.  The
purpose of the present paper is to clarify this issue and to provide a
general survey of soft-Higgs theorems in the framework of renormalized
perturbation theory.

Since it is conceptually simpler, we shall discuss first the
soft-Higgs theorem for pseudoscalar Higgs bosons, which exist in
models with an extended Higgs sector.  In that case the relevant
symmetry is chiral invariance~\cite{GP93}, anomalously broken by the
well-known triangle anomaly~\cite{ABBJ69}.  Next, the theorem
for scalar bosons will be developed, where the anomalies proliferate.
We shall show how they are controlled by the Callan-Symanzik equation,
and give the explicit form of the theorem both in on-shell and minimal
subtraction (MS or \ms) schemes.  The latter allows for the
introduction of effective-theory methods, which already have been
applied in~\cite{D91}.  That the effective-theory picture is
appropriate, follows from the observation that the Higgs coupling to
the heaviest particle (e.g., the top quark) is dominant, and since the
soft-Higgs theorem applies at low energies (and small Higgs masses),
such a particle should be integrated out from the low-energy theory.
Furthermore, when this method is used, logarithms of large mass ratios
are easily summed by the renormalization group.  We shall derive the
form of the soft-Higgs theorem in the effective theory and show that
this framework provides a natural description of all coefficients in
terms of scaling functions, which we shall give in some detail.
In an appendix we give the formulas in a form which is directly
applicable in a next-to-leading order calculation, and demonstrate
their use in a sample calculation that can be compared to the
calculational methods used in the literature.

\section{Pseudoscalars and chiral symmetry}

Before we consider the scalar Higgs, let us investigate the couplings
of a pseudoscalar (CP-odd) Higgs boson, predicted, e.g., by
supersymmetric extensions of the standard model, to fermions.  To keep
things simple, we allow only one external pseudoscalar in the
amplitude.

We apply an infinitesimal global chiral
transformation
\begin{equation}\label{chiral-variation}
  \delta\psi = -i\epsilon g_A\gamma_5\psi/2,
\end{equation}
where $g_A$ is the coupling of the fermion $\psi$ to the pseudoscalar
Higgs $A$:
\begin{equation}
  {\cal L}_A = g_A \bar\psi i\gamma_5 A\psi,
\end{equation}
The variation of the fermion mass term introduces an additional term
in the Lagrangian
\begin{equation}\label{delta-L-chiral}
  \delta{\cal L} = \epsilon g_A m\bar\psi i\gamma_5\psi,
\end{equation}
which may be interpreted as an imaginary contribution to the masses of
left- and right-handed fermions~\cite{GP93}.  To first order in
$\epsilon$, the fermion propagator is modified into
\begin{equation}\label{chiral-propagator}
  G_\epsilon(k) = i\frac{\slash{k} + m + i\epsilon g_A m \gamma_5}{k^2-m^2}.
\end{equation}
Thus we have the relation
\begin{equation}\label{chiral-ward-identity}
  \frac{\partial}{\partial\epsilon}{\cal L}
  = m\frac{\delta{\cal L}_A}{\delta A}
  = \delta{\cal L}.
\end{equation}
In addition to (\ref{delta-L-chiral}), the variation
(\ref{chiral-variation}) introduces terms mixing the scalar ($H$) and
pseudoscalar ($A$) couplings to fermions, since these interactions
also break chiral invariance. This mixing can be compensated by a
redefinition of the $H$ and $A$ fields (in the light-Higgs limit) and
will not be considered here.

Apart from its coupling to fermions, the $A$ field can couple to other
Higgs or Goldstone fields.  The corresponding interaction operator
will be called ${\cal A}$.  This term is not obtained by the
differentiation with respect to $\epsilon$:
\begin{equation}\label{L-chiral-variation}
  \delta{\cal L}
  = m\frac{\delta{\cal L}}{\delta A} - {\cal A}.
\end{equation}
For the full amplitude, differentiation with respect to $\epsilon$
obviously will give no diagrams with ${\cal A}$ insertions.  These
have to be calculated separately, but in the limit of light $A$ they
can be neglected, since the Higgs masses are proportional to
self-couplings.

The relation (\ref{L-chiral-variation}) may be extended to a relation
for $\Gamma$, the generating functional of one-particle
irreducible (1PI) vertices, which is in tree approximation equal to
the action $\int{\cal L}$.  First we write down the identity describing the
irreducible interactions of one zero-momentum $A$ in terms of
zero-momentum operator insertions:
\begin{equation}
  m\int\frac{\delta}{\delta A}\Gamma^{(0)}|_{H=A=0}
  = \int \delta{\cal L}\Downarrow\Gamma^{(0)}|_{H=A=0}
  + \int {\cal A}\Downarrow\Gamma^{(0)}|_{H=A=0}.
\end{equation}
(We use ${\cal A}\Downarrow\Gamma$ as a shorthand for the 1PI
generating functional with an insertion of the renormalized operator
${\cal A}$.)  It holds in the presence of quantum effects, up to
scheme-dependent universal corrections (denoted by $\alpha_A$)
referring to the particular renormalization conditions imposed on the
$A$ field (cf.~Sec.~\ref{sec:scale}):
\begin{equation}
  m(1+\alpha_A)\int\frac{\delta}{\delta A}\Gamma|_{H=A=0}
  = \int \delta{\cal L}\Downarrow\Gamma|_{H=A=0}
  + \int {\cal A}\Downarrow\Gamma|_{H=A=0}.
\end{equation}
This is seen by investigating the diagrams contributing to both sides.

The nontrivial part is the quantum extension of the Ward identity
(\ref{chiral-ward-identity}) of chiral symmetry.  However, the answer
is well known~\cite{ABBJ69}:
\begin{equation}
  \frac{\partial}{\partial\epsilon}\Gamma(\epsilon)|_{\epsilon=0}
  = \int\delta{\cal L}\Downarrow\Gamma
    - s g_A \int \tr{F\tilde F} \Downarrow\Gamma.
\end{equation}
where $s$ is a constant equal to its one-loop value, and $F$ ($\tilde
F$) denotes the (dual) field strength tensor.
Thus we obtain
\begin{equation}\label{SAT}
  m(1+\alpha_A)\int\frac{\delta}{\delta A}\Gamma
  = \frac{\partial}{\partial\epsilon}\Gamma
  + s g_A \int \tr{F\tilde F} \Downarrow\Gamma
  + \int {\cal A}\Downarrow\Gamma,
\end{equation}
where $\epsilon$, $H$, and $A$ have to be set to zero after differentiation.

The relation (\ref{SAT}) is readily verified in an explicit
calculation.  If the integrand contains an open fermion line, the
insertion of a zero-momentum $A$ gives the expression
\begin{equation}
  i\frac{\slash{k} + m}{k^2-m^2}i(g_A i\gamma_5)
  i\frac{\slash{k} + m}{k^2-m^2}
  = i\frac{ig_A\gamma_5}{k^2-m^2}
  = \frac{1}{m}\frac{\partial}{\partial\epsilon}G_\epsilon(k)|_{\epsilon=0},
\end{equation}
where we have used the anticommuting nature of $\gamma_5$.  Thus the
theorem is trivially satisfied.  In fermion loops, the behavior of
$\gamma_5$ is accounted for by the triangle anomaly term in
(\ref{SAT}).

Since we work only to first order in $\epsilon$, it appears only in
the numerator and the differentiation does not affect the
large-momentum behavior.  Thus there is no room for renormalization to
introduce further anomalies.  If we were to derive the interaction of
several pseudoscalar particles, apart from other complications the
quadratic term in $\epsilon$ would give a mass shift, so that
renormalization effects came into play.  Then the scale anomalies
introduced in the next section would have also to be considered.  They
are CP-even and therefore do not occur in the first-order $A$
couplings.

As mentioned in the introduction, the theorem (\ref{SAT}) is valid for
vanishing four-momentum $p_A$.  Only when the mass $m_A$ can be
neglected compared to $m$ --- in which case also the contribution of
${\cal A}$ will be small --- this approaches the amplitude for an
on-shell $A$ particle with $\vec p_A=0$.  For this reason, (\ref{SAT})
may be called a soft-$A$ theorem (for applications, see
ref.~\cite{KS95}, and references cited therein).

\section{Scalars and scale transformations}\label{sec:scale}

Turning over to scalar Higgs couplings in a theory such as the
Standard Model, we divide the Lagrangian
into three parts
\begin{equation}
  {\cal L} = {\cal L}_0 + {\cal L}_{GF} + {\cal L}_H,
\end{equation}
where the first term denotes the Lagrangian of gauge boson and matter
fields, including interactions with the Higgs fields, the second part
is the gauge-fixing and ghost part of the Lagrangian, and the last
term contains the pure Higgs Lagrangian, i.e., the self-interactions
of the physical Higgs $H$ and the Goldstone fields.

We assume that mass terms in ${\cal L}_0$ are generated only through
the Higgs vacuum expectation value $v$, so that
\begin{equation}\label{v-dependence}
  {\cal L}_0(m_i, H) = {\cal L}_0(m_i + g_{i}H)
\end{equation}
where the mass $m_i$ is given by
\begin{equation}
  m_i = g_{i} v.
\end{equation}
Similar to the connection between pseudoscalar couplings and chiral
transformations described in the preceding section, scalar Higgs
couplings are related to scale transformations.  These act on the fields
as
\begin{equation}
  \delta\psi = (d_\psi + x\cdot\partial)\psi,
\end{equation}
where $d_\psi$ is the canonical dimension of the generic field $\psi$.
With $d_\psi=3/2$ for fermions and $d_\psi=1$ for bosons, the
variation of the action reads
\begin{equation}
  \int\delta{\cal L} = \int\left(v\frac{\partial{\cal L}_0}{\partial v}
	+ \delta{\cal L}_{GF} + \delta{\cal L}_H\right),
\end{equation}
since scale invariance is broken by a nonvanishing value of $v$, and
by dimensionful parameters in the gauge-fixing and Higgs parts of the
Lagrangian.  Using (\ref{v-dependence}), this can be rewritten as
\begin{equation}\label{L-variation}
  \int\delta{\cal L} = \int\left(v\frac{\delta{\cal L}}{\delta H}
	- v\frac{\delta{\cal L}_H}{\delta H}
	+ \delta{\cal L}_{GF} + \delta{\cal L}_H \right)
\end{equation}
or
\begin{equation}
  \int\delta{\cal L} = v\int\frac{\delta{\cal L}}{\delta H}
	- \int{\cal H}.
\end{equation}
Similar to the operator ${\cal A}$ in the preceding section, the
operator ${\cal H}$ summarizes Higgs self-couplings and couplings to
gauge fields generated by the gauge fixing.  The diagrams with ${\cal
H}$ insertions have to be calculated explicitly. When they contain
Higgs (Goldstone) self-couplings, in the light-Higgs limit
they can be neglected compared to diagrams with couplings to heavy
particles.  However, the mass parameter in the gauge-fixing part (in a
'tHooft gauge, for instance) introduces a spurious variation
$\delta{\cal L}_{GF}$ that is not related to a Higgs coupling.  In
one-loop order it can be separated just by omitting the derivative
with respect to this gauge-fixing mass in the relations below, but
from the two-loop order on it becomes entangled into the
renormalization procedure.  There are several possibilities to deal
with this complication: one could either use the background-field
method employed in~\cite{KSi95}, which requires the calculation of
more diagrams, or impose the additional restriction that also the
gauge boson masses have to be neglected, or turn over to a
mass-independent renormalization scheme.  The latter will be done in
the next section.

The extension of (\ref{L-variation}) to irreducible vertices has the form
\begin{equation}\label{Gamma-variation}
  (1+\alpha_H)v\int\frac{\delta}{\delta H}\Gamma
	= \int\delta{\cal L}\Downarrow\Gamma
	  + \int{\cal H}\Downarrow\Gamma.
\end{equation}
The unrenormalized diagrams which do not involve Higgs (Goldstone)
self-couplings, contributing to the relation (\ref{Gamma-variation}),
are identical on both sides.  The same is true for the counterterms,
except for the renormalization of the external Higgs field: In the
on-shell scheme the left-hand side is renormalized with the momentum
$p_H$ flowing into the vertex satisfying $p_H^2=m_H^2$, whereas the
right-hand side is defined for vanishing momentum $p_H$.  Corrections
proportial to $m_H^2$ are consistently absorbed into ${\cal H}$, so
that there remains an overall multiplicative factor, denoted by
$1+\alpha_H$.  Any possible correction involving Higgs (Goldstone)
self-couplings is also absorbed in~${\cal H}$.

We can employ the Ward identity of scale invariance to get rid of the
operator $\delta{\cal L}$ in (\ref{Gamma-variation}) and obtain in
this way a nontrivial statement.  Although scale invariance is broken
by quantum corrections~\cite{Col71}, the quantum action
principles~\cite{QAP} tell us that all corrections (anomalies) are
given by a linear combination of local operator insertions with the
appropriate quantum numbers and dimension.  The resulting
anomalous Ward identity of scale transformations is known as the
Callan-Symanzik equation~\cite{CS70}.  For an abelian Higgs
model a detailed derivation is given in~\cite{KSi95}.  Here we simply
generalize the result for a general Higgs model in an on-shell
scheme:  The relation (\ref{Gamma-variation}) can be replaced by
\begin{eqnarray}\label{SHT}
  (1+\alpha_H)v\int\frac{\delta}{\delta H}\Gamma^{(n_1,\ldots)}
	&=&
	\left(\sum_i m_i\frac{\partial}{\partial m_i}
		+ \beta_i\frac{\partial}{\partial g_i}
		+ \frac{n_i}{2}\gamma_i
		+ \delta_i\frac{\partial}{\partial\xi_i}\right)
	\Gamma^{(n_1,\ldots)} \nonumber\\
	&&
	+\, \int{\cal H}\Downarrow \Gamma^{(n_1,\ldots)}.
\end{eqnarray}
for an irreducible vertex with $n_i$ external fields of species $i$,
which is given by
\begin{equation}
  \Gamma^{(n_1,\ldots)}
	= \frac{\delta^{n_1}}{\delta\psi_1^{n_1}}\cdots \Gamma.
\end{equation}
The $\xi_i$ are gauge parameters, and ${\cal H}$ can be neglected, as
discussed above.

The coefficients $\alpha$, $\beta$, $\gamma$, $\delta$ have to be
determined order by order in perturbation theory.  They are universal
for all processes, but they are mass-dependent and are not simply
related to the familiar scaling functions in a mass-independent
scheme, which will be introduced in the next section.  Some of them
are determined by the symmetry.  In particular, all but one of the
coupling constants are usually expressed in terms of masses, so that
their beta functions may be identified as mass beta
functions~\cite{KSi95}, due to the normalization conditions one
imposes.

Introducing the concept of bare parameters, which can also be used for
a simple derivation of the Callan-Symanzik equation~\cite{Col71}, the
coefficients can be expressed as derivatives of the renormalized
parameters with respect to bare masses.  The relation~(\ref{SHT}) thus
involves the operations carried out in~\cite{KS94a,KS94b} in reverse
order: If instead of renormalizing after taking the derivative with
respect to bare masses, the renormalized amplitudes are
differentiated, one has to correct for the derivatives of the
renormalized parameters.  These are the corrections summarized
in~(\ref{SHT}).

\section{Renormalization group}

The relation (\ref{SHT}) has been derived within the context of an
on-shell scheme, where all dimensionful parameters are expressed in
terms of the physical masses of the theory.  When a minimal
subtraction scheme (MS or \ms) is used, the derivation is greatly
simplified.  The price are complicated, but calculable, relations of
the parameters to physical observables.

Taken literally, a minimal subtraction scheme does not subtract the
tadpole diagrams completely.  This is a minor problem:  In terms of
irreducible diagrams, only the Higgs one-point function is affected,
and since it is a constant, the finite part can completely be
subtracted without affecting properties of the dimensional
renormalization procedure that follow from mass independence: tadpoles
are simply omitted.

To lowest order (tree level), the relation (\ref{SHT}) reads
\begin{equation}\label{SHT-tree}
  v\int\frac{\delta}{\delta H}\Gamma^{(n_1,\ldots)}
	=
	\sum_i m_i\frac{\partial}{\partial m_i}
	\Gamma^{(n_1,\ldots)} \nonumber\\
	+ \int{\cal H}\Downarrow \Gamma^{(n_1,\ldots)}.
\end{equation}
Here we exclude the mass term in the gauge fixing from the
derivative, so that ${\cal H}$ does not contain spurious couplings and
is truly negligible in the light-Higgs limit.

In the dimensional renormalization scheme, the quantum action
principles hold in the strong sense~\cite{BM77}, so that
(\ref{SHT-tree}) is valid without corrections even after
renormalization, if the running parameters at the scale $\mu$ are
inserted everywhere.  Stated differently, the coefficients $\beta$ and
$\gamma$ appearing in (\ref{SHT}), which are related to mass
derivatives of the renormalized parameters, vanish identically: the
renormalization factors are mass-independent.

It is now easy to get rid of the mass derivatives in favor of familiar
renormalization group coefficients, if we have a one-scale problem.
The case of multiple scales will be considered in the next sction.
We thus put all other masses and external momenta to zero, ignoring
infrared divergences.  (If some care in the renormalization of
subdivergences is taken, they may be regulated dimensionally.)  We use
the MS or \ms\ scheme.  Then any renormalized vertex function with
mass dimension $d$ and $n_\ell$ external light fields of species
$\ell$ can be expressed as
\begin{equation}
  \Gamma(m(\mu),\mu) = m(\mu)^d
	\tilde \Gamma(\ln{\textstyle\frac{\mu}{m(\mu)}}, g(\mu)),
\end{equation}
with a dimensionless function $\tilde\Gamma$.  Defining the mass anomalous
dimension
\begin{equation}
  \frac{d\ln m(\mu)}{d\ln\mu} = -\gamma_m(\mu),
\end{equation}
we derive
\begin{equation}
  \frac{d}{d\ln\mu}\Gamma(\mu)
	= \left(d - (1+\gamma_m)\frac{\partial}{\partial\ln m(\mu)}
	  + \beta_i\frac{\partial}{\partial g_i}\right)
	  \Gamma(m(\mu),\mu).
\end{equation}
On the other hand, the vertex function satisfies a renormalization
group equation
\begin{equation}
  \frac{d}{d\ln\mu}\Gamma(\mu)
	= -\frac{n_\ell}{2}\gamma_\ell \Gamma(\mu),
\end{equation}
so that
\begin{equation}
  \frac{\partial}{\partial\ln m(\mu)}\Gamma(m(\mu),\mu)
	= \frac{1}{1+\gamma_m}\left(
	\beta_i\frac{\partial}{\partial g_i}
	+ d + \frac{n_\ell}{2}\gamma_\ell\right)
	\Gamma(m(\mu),\mu).
\end{equation}
The definitions of the scaling coefficients are
\begin{eqnarray}\label{beta-gamma-MS}
  \beta_i &=& \frac{d}{d\ln\mu}g_i,\\
  \gamma_i &=& \frac{d}{d\ln\mu}\ln Z_i.
\end{eqnarray}
Thus the soft-Higgs theorem in the dimensional renormalization scheme
has the form
\begin{eqnarray}\label{SHT-MS}
  v\int\frac{\delta}{\delta H}\Gamma^{(n_1,\ldots)}
	&=& \frac{1}{1+\gamma_m}
	\left(\beta_i\frac{\partial}{\partial g_i}
		+ d + \frac{n_i}{2}\gamma_i \right)
	\Gamma^{(n_1,\ldots)} \nonumber\\
	&&
	+\, \int{\cal H}\Downarrow \Gamma^{(n_1,\ldots)}.
\end{eqnarray}
This looks very similar to (\ref{SHT}), but without the derivative
with respect to the mass: In this scheme the coupling of the
Higgs is given by scaling coefficients only, for vanishing external
momenta.

\section{Heavy and light fields: effective theory}

The arguments leading to (\ref{SHT-MS}) are sufficient in a
single-scale problem, when the scale $\mu$ can be chosen not much
different from $m$, and infrared divergences can be ignored.  However,
in particular when QCD corrections are considered, it is customary to
resum logarithms because of the comparatively large coupling constant.
In this situation it is mandatory to apply the method of effective
field theory~\cite{Georgi}, since below a
mass threshold the renormalization group of a mass-independent scheme
is not able to resum logarithms correctly (see~\cite{Kr94} for a
discussion of this point).

This framework is also helpful to understand the role
of different scales in the presence of light masses and small momenta.
It clearly separates the infrared behaviour from the ultraviolet, and
all coefficients in the soft-Higgs theorem can be expressed in terms
of anomalous dimensions and beta functions.

The heavy particle is integrated out at a scale $\mu_0$ of the order
of its mass $m_h$, with the effect that the Lagrangian ${\cal L}$
which can be divided into a part ${\cal L}_\ell$ that contains only
light fields, and the rest ${\cal L}_h$
\begin{equation}
  {\cal L} = {\cal L}_\ell + {\cal L}_h,
\end{equation}
is replaced by an effective Lagrangian
\begin{equation}\label{L-eff}
  \hat{\cal L} = {\cal L}_\ell + \sum_i \hat C_i {\cal O}_i.
\end{equation}
The operators ${\cal O}_i$ are of increasing dimension, divided by
appropriate powers of $m_h$.  They consist only of light fields.
In particular, they contain operators of dimension four or less which
are already present in ${\cal L}_\ell$.  It is convenient to absorb
those into a redefinition of parameters:
\begin{equation}\label{L-eff'}
  \hat{\cal L} = \hat{\cal L}_\ell + \sum_{{\rm dim}>4} \hat C_i {\cal O}_i.
\end{equation}
The ordering of operators according to their dimension is appropriate
when no light masses are around.  Otherwise it simplifies the
discussion if the series is organized in terms of powers of $1/m_h$
instead.

One should keep in mind that the limit under consideration is not
exactly a heavy-mass limit for, e.g., the top quark ($m_t\to\infty$
would imply strong coupling), but rather a small-coupling limit for
the Higgs self-coupling.  The use of effective field theory methods in
this limit for the resummation of logarithms is somewhat unusual, but
in perturbation theory where the mass of a particle and its coupling
are clearly separated, there is no real difference to an ordinary
theory with large mass ratios.  In particular, the fact that a
particle like the top quark may be of a non-decoupling nature
introduces no practical difficulties, since we stay in the
weak-coupling regime where the mass is still small compared to the
strong-coupling scale $4\pi v$.

The coefficients $\hat C_i$ are determined as power series in the
couplings by a matching calculation.  It involves calculating the
difference of diagrams in the effective and full theories, where the
mass has to be kept in the full-theory diagrams.  With respect to the
heavy particle it thus incorporates the change from a mass-independent
to essentially an on-shell scheme (for details, see
e.g.~\cite{Collins}).  This is necessary to ensure that below
threshold the correct logarithms are summed.

An important property of the matching coefficients $\hat C_i$ in
(\ref{L-eff}) is that they are infrared safe quantities: They are
infrared convergent, and they contain no logarithms of light
parameters $m_\ell$.  Thus any dimensionless matching contribution
$\Delta$ is a function of the ratio of the two dimensionful parameters
$m_h(\mu_0)$ and $\mu_0$, which appear only logarithmically, and of
the dimensionless couplings $g(\mu_0)$:
\begin{equation}
  \Delta = \Delta({\textstyle \ln\frac{\mu_0}{m_h(\mu_0)}},g(\mu_0)).
\end{equation}
It is independent of light masses and momenta, up to power corrections
$O(m_\ell/m_h,p_\ell/m_h)$ which are absorbed in the higher-order
terms.

Since masses are renormalized multiplicatively, the vector
$\partial/\partial m(\mu)$ scales contravariant to the mass vector
$m(\mu)$, and the scalar product which appears in (\ref{SHT-tree}) is
invariant:
\begin{equation}
  \sum_i m_i(\mu)\frac{\partial}{\partial m_i(\mu)}
  = \sum_i m_i(\mu_0)\frac{\partial}{\partial m_i(\mu_0)}.
\end{equation}
Thus we can take derivatives at the matching scale where the heavy
particle is removed from the theory.  In the effective theory, the
heavy mass $m_h(\mu_0)$ appears implicitly in the coefficients,
whereas the light masses $m_\ell(\mu_0)$ (reexpressed in terms of
effective masses $\hat m_\ell(\mu)$) remain dynamical parameters.

\section{Mass dependence of effective parameters}

Before we state the soft-Higgs theorem in the effective theory, we
first have to calculate the dependence of the parameters in the
low-energy effective theory on the heavy mass $m_h$.  Let us consider
first the running coupling constants.  In the full theory, they
satisfy
\begin{equation}\label{alpha-running}
  \frac{d}{d\ln\mu}g(\mu) = \beta(g(\mu))
\end{equation}
where $g$ and $\beta$ are vectors so that this is a coupled
system of differential equations in the general case.  By definition,
in a mass-independent scheme such as \ms\ the value of a running
coupling constant does not depend on masses
\begin{equation}
  \frac{\partial}{\partial\ln m_h}g(\mu) = 0,
\end{equation}
although the existence of the heavy particle affects the beta
function.

The transition from (\ref{L-eff}) to (\ref{L-eff'}) is reflected in
the matching condition
\begin{equation}\label{alpha-matching}
  \hat g_0 \equiv \hat g(\mu_0) = g(\mu_0)
	+ \Delta g({\textstyle\ln\frac{\mu_0}{m_h(\mu_0)}}, g(\mu_0)),
\end{equation}
where $\Delta g$ is a polynomial function of $g$ starting with
the cubic term.  Only diagrams containing the heavy particle
contribute to $\Delta g$.  In the effective theory, the running
coupling constants satisfy
\begin{equation}\label{alpha-eff-running}
  \frac{d}{d\ln\mu}\hat g(\mu) = \hat\beta(\hat g(\mu))
\end{equation}
where $\hat\beta$ is obtained from $\beta$ by omitting the diagrams
containing the heavy particle.  The renormalization group can be used
to resum logarithms and to obtain the value of $\hat g$ at another
scale $\mu$
\begin{equation}\label{alpha-resummed}
  \hat g(\mu)
	= \hat g({\textstyle\ln\frac{\mu}{\mu_0}}, \hat g_0).
\end{equation}
When evaluated exactly, $\hat g(\mu)$ is in fact independent of
$\mu_0$:
\begin{equation}
  \frac{\partial\hat g(\mu)}{\partial\ln\mu_0} = 0.
\end{equation}
Inserting the definitions (\ref{alpha-resummed}) and
(\ref{alpha-matching}) into this identity, and using
(\ref{alpha-running}) and (\ref{alpha-eff-running}), we derive the
relation
\begin{eqnarray}\label{dg-dm}
  \frac{\partial\hat g(\mu)}{\partial\ln m_h(\mu_0)}
	&=& \frac{1}{1+\gamma_{hh}(\mu_0)+
	\frac{m_\ell}{m_h}\gamma_{\ell h}(\mu_0)}
	\nonumber\\
	&&\times\left[
	\frac{\partial\hat g(\mu)}{\partial\hat g_0}
	\left(1 + \frac{\partial\,\Delta g(\mu_0)}{\partial g(\mu_0)}
	\right)\beta(\mu_0) - \hat\beta(\mu)\right].
\end{eqnarray}

The other parameters of the effective theory include the matrix of
field renormalization factors $\hat\zeta$ (with $\hat Z =
\hat\zeta^T\hat\zeta$ being the coefficient of the kinetic term in the
effective Lagrangian), the masses $\hat m_\ell$ of light fields (we
consider only fermions), and the coefficients $\hat C_k$ of additional
operators of the order $1/m_h^k$.  In the full theory, the
renormalization group equations are
\begin{eqnarray}
  \frac{d\zeta}{d\ln\mu} &=& \frac12\zeta(\mu)\,\gamma(\mu),\\
  \frac{dm_\ell}{d\ln\mu} &=&
	-m_h(\mu)\,\gamma_{h\ell}(\mu) - m_\ell(\mu)\,\gamma_{\ell\ell}(\mu),
	\\
  \frac{dm_h}{d\ln\mu} &=&
	-m_h(\mu)\,\gamma_{hh}(\mu) - m_\ell(\mu)\,\gamma_{\ell h}(\mu).
\end{eqnarray}
The matching conditions yield
\begin{eqnarray}
  \hat\zeta(\mu_0) &=& \bar\zeta(\mu_0)
	\left[1 + \Delta\zeta({\textstyle\ln\frac{\mu_0}{m_h(\mu_0)}},
		g(\mu_0))\right], \\
  \hat m_\ell(\mu_0) &=& m_h(\mu_0)\,
	\Delta\sigma({\textstyle\ln\frac{\mu_0}{m_h(\mu_0)}}, g(\mu_0))
	\nonumber\\
	&& +\, m_\ell(\mu_0)\left[
	1 + \Delta\tau({\textstyle\ln\frac{\mu_0}{m_h(\mu_0)}},
		g(\mu_0)) \right], \\
  \hat C_k(\mu_0) &=& m_h(\mu_0)^{-k}
	\Delta C_k({\textstyle\ln\frac{\mu_0}{m_h(\mu_0)}}, g(\mu_0)),
\end{eqnarray}
where $\bar\zeta$ is equal to $\zeta$ projected onto the space of
light fields.

In the effective theory, the coefficients satisfy renormalization
group equations
\begin{eqnarray}
  \frac{d\hat\zeta}{d\ln\mu} &=&
	\frac12\hat\zeta(\mu)\,\hat\gamma(\mu),\\
  \frac{d\hat m_\ell}{d\ln\mu} &=&
	- \hat m_\ell(\mu)\,\hat\gamma_{\ell\ell}(\mu),
\end{eqnarray}
and
\begin{equation}\label{C-running}
  \left(\frac{d}{d\ln\mu} + \hat\gamma^T_k\right)
	\hat C_k(\mu)
	+ \sum_{\Sigma\ell=k}\hat\gamma_{\ell_1\ell_2\cdots}^T
	  \hat C_{\ell_1}(\mu)\,\hat C_{\ell_2}(\mu)\cdots = 0.
\end{equation}
By definition, operators mix among each other only if they have the
same power $m_h^{-k}$ as prefactor.  The nonlinear terms appear for
$k>1$; they originate from time-ordered products of lower-dimensional
operators mixing into the local operators.  The renormalization group
equations can be solved iteratively for increasing dimension so that
the nonlinear terms play the role of the driving term in an
inhomogeneous differential equation for the coefficients with index
$k$.

The solution of the renormalization group equations can be cast into
the form
\begin{eqnarray}
  \hat\zeta(\mu)
	&=& \hat\zeta(\mu_0)\,
	\hat R({\textstyle\ln\frac{\mu}{\mu_0}},g(\mu_0)),\\
  \hat m_\ell(\mu)
	&=& \hat m_\ell(\mu_0)\,
	\hat R_{\ell\ell}({\textstyle\ln\frac{\mu}{\mu_0}},g(\mu_0)),\\
  \hat C_k(\mu)
	&=&
	\hat R_k^T({\textstyle\ln\frac{\mu}{\mu_0}},g(\mu_0))\,
	\hat C_k(\mu_0) \nonumber\\
	&& + \sum_{\Sigma\ell=k}
	\hat R_{\ell_1\ell_2\cdots}^T
		({\textstyle\ln\frac{\mu}{\mu_0}},g(\mu_0))
	\,\hat C_{\ell_1}(\mu_0)\,\hat C_{\ell_2}(\mu_0)\cdots.
\end{eqnarray}
Requiring $\mu_0$ independence of these expressions, we obtain
\begin{eqnarray}
  \hat\zeta^{-1}\frac{\partial\hat\zeta}{\partial\ln m_h}
	&=& \frac{1}{1+\gamma_{hh}+\frac{m_\ell}{m_h}\gamma_{\ell h}}
	\left\{
	\vphantom{\frac{\partial}{\partial g}}
	\hat R^{-1}\left(1+\Delta\zeta\right)^{-1}
	\frac12\bar\gamma\left(1+\Delta\zeta\right)\hat R
	- \frac12\hat\gamma
	\right.\nonumber\\
	&& \quad\quad \left.
	+\,\hat R^{-1}
	\left(1+\Delta\zeta\right)^{-1}
	\beta\frac{\partial}{\partial g}
	\left[\left(1+\Delta\zeta\right)\hat R\right]\right\},
	\label{dzeta-dm}\\
  \frac{\partial\hat m_\ell}{\partial\ln m_h}
	&=& \frac{1}{1+\gamma_{hh}+\frac{m_\ell}{m_h}\gamma_{\ell h}}
	\left\{
	m_h\left[ \Delta\sigma\,\hat R_{\ell\ell}
	+ \beta\frac{\partial}{\partial g}
	\left(\Delta\sigma\,\hat R_{\ell\ell}\right)
	\right.\right.\nonumber\\
	&&\quad\quad\quad\quad\left.
	\vphantom{\frac{\partial}{\partial g}}
	- \gamma_{h\ell}(1+\Delta\tau)\hat R_{\ell\ell}
	+ \Delta\sigma\,\hat R_{\ell\ell}\hat\gamma_{\ell\ell}
	\right]
	\nonumber\\
	&&\quad\quad
	+\, m_\ell \left[
	\beta\frac{\partial}{\partial g}
	\left((1+\Delta\tau)\hat R_{\ell\ell}\right)
	\right.\nonumber\\
	&&\quad\quad\quad\quad\left.\left.
	\vphantom{\frac{\partial}{\partial g}}
	- \gamma_{\ell\ell}(1+\Delta\tau)\hat R_{\ell\ell}
	+ (1+\Delta\tau)\hat R_{\ell\ell}\hat\gamma_{\ell\ell}
	\right]
	\right\}
	\label{dm-dm}\\
  \frac{\partial\hat C_k}{\partial\ln m_h}
	&=& \frac{1}{1+\gamma_{hh}+\frac{m_\ell}{m_h}\gamma_{\ell h}}
	\left\{
	\left(-k+\hat\gamma_k^T + \beta\frac{\partial}{\partial g}\right)
	\hat C_k
	\right.\nonumber\\
	&& \quad\quad \left.
	\vphantom{\frac{\partial}{\partial g}}
	+\, \sum_{\Sigma\ell=k}\hat\gamma_{\ell_1\ell_2\cdots}^T
	  \hat C_{\ell_1}\hat C_{\ell_2}\cdots
	\right\}.
	\label{dC-dm}
\end{eqnarray}
In the first equation, $\bar\gamma$ is the anomalous dimension matrix
$\gamma$ projected onto the space of light fields.  All full-theory
and matching coefficients are evaluated at the scale $\mu_0$, and the
effective theory coefficients (denoted by a hat) at the scale $\mu$.
It is convenient to choose
\begin{equation}
  \mu_0 = m_h(\mu_0) = m_h(m_h),
\end{equation}
so that all logarithms vanish in the matching coefficients.

When in the effective theory the logarithms have been resummed, the
solutions $\hat R_x(\mu)$ are usually available as functions
\begin{equation}
  \hat R_x(\mu) = \hat R_x(\hat g_0,\hat g(\mu)),
\end{equation}
with $\hat g_0$ from (\ref{alpha-matching}), containing no explicit
logarithms.  Then the derivative with respect to $g(\mu_0)$ is given
by
\begin{eqnarray}
  \frac{\partial\hat R_x}{\partial g(\mu_0)}
  &=& \beta(\mu_0)
	\left(1 + \frac{\partial\,\Delta g}{\partial g(\mu_0)}\right)
	\frac{\partial\hat R_x}{\partial\hat g_0}
	\\
  	&& +\,\left[ \beta(\mu_0)
	\left(1 + \frac{\partial\,\Delta g}{\partial g(\mu_0)}\right)
	\frac{\partial\hat g(\mu)}{\partial\hat g_0}
	- \hat\beta(\mu)\right]
	\frac{\partial\hat R_x}{\partial\hat g(\mu)}.
	\nonumber
\end{eqnarray}

\section{Soft-Higgs theorem in the effective theory}

We are now in a position to derive the soft-Higgs theorem for the
effective theory.  Applying the chain rule to (\ref{SHT-tree}), we
obtain
\begin{eqnarray}\label{SHT-MS-full}
  v\int\frac{\delta}{\delta H}\Gamma^{(n_1,\ldots)}
	&=& \sum_\ell\frac{1}{1+\frac{m_h\Delta\sigma}{m_\ell(1+\Delta\tau)}}
	\hat m_\ell\frac{\partial}{\partial\hat m_\ell}
	\hat\Gamma^{(n_1,\ldots)}
	\nonumber\\
	&& +\left(
	\frac{d\hat m_\ell}{d\ln m_h}
		\frac{\partial}{\partial\hat m_\ell}
	+ \frac{d\hat g}{d\ln m_h}
		\frac{\partial}{\partial\hat g}
	+ n_i\zeta_i^{-1}
		\frac{d\zeta_i}{d\ln m_h}
	\right)\hat\Gamma^{(n_1,\ldots)}
	\nonumber\\
	&&+\int\left( \frac{d\hat C_k}{d\ln m_h}
		{\cal O}_k
	+ {\cal H}\right)\Downarrow\hat\Gamma^{(n_1,\ldots)},
\end{eqnarray}
with $\hat\zeta_\ell=(\hat Z_\ell)^{1/2}$ being the normalization of
the light particle $\ell$ which appears $n_\ell$ times in the vertex
function. (We have neglected the mixing in the light sector for
simplicity.)  It is understood that Yukawa couplings are expressed in
terms of masses only after the differentiations have been carried out.

This equation replaces (\ref{SHT-MS}) in the general case, taking now
full account of light masses and nonvanishing momenta.  When the
explicit expressions (\ref{dg-dm}), (\ref{dzeta-dm}), (\ref{dm-dm}),
and (\ref{dC-dm}) are inserted, a simple pattern emerges: The
coefficients essentially consist of differences of scaling functions
(beta functions and anomalous dimensions) in the full and effective
theories.  In addition, beginning from second order the mass
dependence hidden in the matching contributions has to be accounted
for.  (In the non-decoupling case, where heavy and light masses are
allowed to mix, a matching contribution enters already at leading
order.)  This generalizes the observation in~\cite{VVSS79} that the
leading contribution to, e.g., the $H\to\gamma\gamma$ decay amplitude,
is determined by the contribution of heavy particles to the beta
function, which is equal to the difference of beta functions in the
full and effective theories.

The parameter set of the effective theory (denoted by a hat) is
reduced.  Along with the heavy mass it does no longer contain the
couplings of the heavy particle.  Thus there are no problems arising
from the fact that its Higgs coupling is related to its mass and
should be defined at the same scale.  If we were to use the full
renormalization group in a mass-dependent scheme (to account for mass
effects), with a dynamical heavy particle below threshold, we would
run into difficulties~\cite{Kr93}.

\section{Conclusions}

The soft-Higgs theorem has found a wide range of applications in
standard model calculations, for Higgs amplitudes in the limit of
small Higgs mass and momentum, where it can be used for approximations
and as a nontrivial check for complete calculations.  The aim of this
paper was to clarify its meaning in the context of renormalized
perturbation theory.  We have shown that there is a close connection
to the Ward identity of scale invariance, the Callan-Symanzik
equation.  Similarly, broken chiral symmetry is related to the
coupling of pseudoscalar Higgs fields.

The exact form of the renormalized soft-Higgs theorem depends on the
renormalization scheme one has imposed.  We presented results in the
on-shell scheme, and in a minimal subtraction scheme (MS or \ms),
where things simplify, and in particular the coefficients are mass
independent.  If large mass ratios are present, so that logarithms
have to be summed up, heavy particles have to be integrated out below
threshold, and the soft-Higgs theorem assumes a different form.  We
have calculated the coefficients in a fairly general way, so that it
should be straightforward to use them in a particular problem.  The
many possible extensions of the standard model leave plenty of room
for new particles and interactions up to the TeV range, where the
soft-Higgs theorem could be valuable in the calculation of Higgs
interactions.

\subsection*{Acknowledgements}
I am grateful to B.~A.~Kniehl, M.~Kr\"amer, E.~Kraus, T.~Mannel,
T.~Ohl, and M.~Spira for enlightening discussions.  In addition, I
thank E.~Kraus for drawing my attention to ref.~\cite{KSi95}, M.~Spira
for providing me a copy of ref.~\cite{KS95} prior to publication, and
T.~Ohl for reading and useful comments on the manuscript.  Special
thanks I would like to express to B.~A.~Kniehl for the invitation and
the atmosphere during the 1995 Ringberg workshop on electroweak
interactions.

\section*{Appendix: NLO expansion}

To establish the connection to applications of the soft-Higgs theorem
that have been considered in the literature, we expand the various
terms up to next-to-leading order in the coupling constants.  The
complicated structure of the full standard model, including QCD,
forces us to maintain full generality and to keep the mixing of the
various coupling constants, fields, and masses.  However, there may be
no need to resum logarithms, as it has been the case in existing
applications (for instance, with the present knowledge of Higgs and
top masses, $\ln(m_t/m_H)$ is not particularly large).  For this
reason, we neglect higher-order logarithmic terms in the formulas
below and express everything in terms of $g^i=g^i(\mu_0)$,
$\zeta=\zeta(\mu_0)$, and $m_\ell=m_\ell(\mu_0)$, where the matching
point $\mu_0$ is chosen equal to $m_h(\mu_0)$.  With
$g^{ij\cdots}\equiv g^i g^j\cdots$, and a summation convention for
upper indices understood (all tensors may be chosen symmetrically), we
expand up to next-to-leading order:
\begin{eqnarray}
  \hat g^i(\mu)
	&=& g^i
		+ g^{ijk}\left(\hat\beta_{i1}^{jk}\ln\frac{\mu}{ m_h}
		+ \Delta g_{i1}^{jk}\right), \\
  \hat\zeta(\mu)
	&=& \zeta\left(1
		+ g^{ij}\Delta\zeta_1^{ij}\right) \hat R(\mu),  \\
  \hat m_\ell(\mu)
	&=& \left[m_\ell\left(1 + g^{ij}\Delta\tau_1^{ij}\right)
		+  m_h\left(g^{ij}\Delta\sigma_1^{ij}
			+ g^{ijkl}\Delta\sigma_2^{ijkl}\right)\right]
		\hat R_{\ell\ell}(\mu), \\
  \hat C_k(\mu)
	&=& \frac{1}{ m_h^k}\left[1
		- g^{ij}\hat\gamma_{k1}^T\ln\frac{\mu}{ m_h}\right]
		g^{lm\cdots}\Delta C_{k1}^{lm\cdots}.
\end{eqnarray}
$\hat R(\mu)$ and $\hat R_{\ell\ell}(\mu)$ are chosen to be solutions of the
renormalization group equation valid to order $g^4$.  The initial
Wilson coefficients $\Delta C_k$ are assumed to be of order $g^n$,
with $n\geq 2$.  The expansions of the scaling functions are
\begin{eqnarray}
  \hat\beta(g) &=& g^{ijk}\hat\beta_{i1}^{jk}
	+ g^{ijklm}\hat\beta_{i2}^{jklm}, \\
  \hat\gamma(g) &=& g^{ij}\hat\gamma_1^{ij}
	+ g^{ijkl}\hat\gamma_2^{ijkl},
\end{eqnarray}
and so on.  A straightforward calculation then gives
\begin{eqnarray}\label{derivative-expansions}
  \label{coupling-expansion}
  \frac{d\hat g^i}{d\ln m_h}
	&=& g^{ijk}\left(\beta_{i1}^{jk} - \hat\beta_{i1}^{jk}\right)
	+ g^{ijklm}\left(\beta_{i2}^{jklm}-\hat\beta_{i2}^{jklm}\right)
	\\
	&& +\, g^{ijklm}\left(\beta_{i1}^{jk} - \hat\beta_{i1}^{jk}\right)
	                \left[\hat\beta_{i1}^{lm}\ln\frac{\mu}{ m_h}
		              + \Delta g_{i1}^{lm} - \gamma_{hh1}^{lm}\right]
	\nonumber\\
	&& +\, 2g^{ijklm}
                \left[ \hat\beta_{i1}^{jk}
	               \left(\beta_{j1}^{lm} - \hat\beta_{j1}^{lm}\right)
	               \ln\frac{\mu}{ m_h}
	             + \Delta g_{i1}^{jk}\beta_{j1}^{lm}
	             - \hat\beta_{i1}^{jk}\Delta g_{j1}^{lm}\right],
	\nonumber\\
  \label{zeta-expansion}
  \hat\zeta^{-1}\frac{d\hat\zeta}{d\ln m_h}
	&=&  g^{ij}
	\frac12 \left(\gamma_{1}^{ij}-\hat\gamma_{1}^{ij}\right)
	\\
	&& +\, g^{ijkl}\left[
        \frac12 \left(\gamma_{2}^{ijkl}-\hat\gamma_{2}^{ijkl}\right)
        - \frac12 \gamma_{hh1}^{ij}
	        \left(\gamma_{1}^{kl}-\hat\gamma_{1}^{kl}\right)
        \right.	\nonumber\\
	&& \quad\quad
        +\,\hat\gamma_{ 1}^{ij}
	        \left(\beta_{i1}^{kl}-\hat\beta_{i1}^{kl}\right)
	        \ln\frac{\mu}{ m_h}
	+ 2\Delta\zeta_1^{ij}\beta_{i1}^{kl}
	- \hat\gamma_{ 1}^{ij}\Delta g_{i1}^{kl}
	\nonumber\\
	&& \quad\quad \left.
	+\, \frac12 [\gamma_{ 1}^{ij}, \Delta\zeta_1^{kl}]
	+ \frac14 [\gamma_{ 1}^{ij}, \hat\gamma_{ 1}^{kl}]
	          \ln\frac{\mu}{ m_h} \right],
	\nonumber\\
  \frac{d\hat m_\ell}{d\ln m_h}
	&=&  m_h\left\{  g^{ij}
                \left(\Delta\sigma_1^{ij}-\gamma_{h\ell1}^{ij}\right)
		\right.
	\\
	&& \quad +\,g^{ijkl}\left[
                \left(\Delta\sigma_2^{ijkl}-\gamma_{h\ell2}^{ijkl}\right)
		\right.
	\nonumber\\
	&& \quad\quad\quad
		-\left(\Delta\sigma_1^{ij}-\gamma_{h\ell1}^{ij}\right)
		       \hat\gamma_{\ell\ell 1}^{kl}\ln\frac{\mu}{m_h}
		- \gamma_{hh1}^{ij}
		       \left(\Delta\sigma_1^{kl}-\gamma_{h\ell1}^{kl}\right)
	\nonumber\\
	&& \quad\quad\quad \left.\left.
		+\,\Delta\sigma_1^{ij}\hat\gamma_{\ell\ell1}^{kl}
		- \gamma_{h\ell 1}^{ij}\Delta\tau_1^{kl}
		+ 2\Delta\sigma_1^{ij}\beta_{i1}^{kl}\right]
		\right\}
	\nonumber\\
	&& +\,m_\ell\left\{   \vphantom{\frac{\mu}{m_h}}
		g^{ij}
                \left(\hat\gamma_{\ell\ell 1}^{ij}
                      -\gamma_{\ell\ell 1}^{ij}\right)
	\right.
	\nonumber\\
	&& \quad
		+\,g^{ijkl}
                \left[\left(\hat\gamma_{\ell\ell 2}^{ijkl}
                            -\gamma_{\ell\ell 2}^{ijkl}\right)
                    - \gamma_{hh1}^{ij}
	              \left(\hat\gamma_{\ell\ell 1}^{kl}
                            -\gamma_{\ell\ell 1}^{kl}\right)
		\vphantom{\frac{\mu}{m_h}}\right.
	\nonumber\\
	&& \quad\quad\quad
		-\,\left(\hat\gamma_{\ell\ell 1}^{ij}
			 - \gamma_{\ell\ell 1}^{ij}\right)
		        \hat\gamma_{\ell\ell 1}^{kl}\ln\frac{\mu}{m_h}
		-\,\gamma_{\ell h1}^{ij}\left(\Delta\sigma_1^{kl}
			- \gamma_{h\ell 1}^{kl}\right)
	\nonumber\\
	&& \quad\quad\quad
		+\,\Delta\tau_1^{ij}\hat\gamma_{\ell\ell1}^{kl}
		- \gamma_{\ell\ell1}^{ij}\Delta\tau_1^{kl}
		+ 2\Delta\tau_1^{ij}\beta_{i1}^{kl}
	\nonumber\\
	&& \quad\quad\quad \left.\left.
		+\,2\hat\gamma_{\ell\ell1}^{ij}
		        \left(\hat\beta_{i1}^{kl}-\beta_{i1}^{kl}\right)
		        \ln\frac{\mu}{m_h}
		+ 2\hat\gamma_{\ell\ell1}^{ij} \Delta g_{i1}^{kl}
		\right]\right\}
	\nonumber\\
  \frac{d\hat C_k}{d\ln m_h}
	&=&  \frac{1}{ m_h^k}
	\left\{ -k g^{lm\cdots}
	        - g^{ijlm\cdots}(\hat\gamma_{k1}^T)^{ij}
	        + ng^{ijlm\cdots}\beta_{l1}^{ij} \right\}
	C_{k1}^{lm\cdots}.
\end{eqnarray}
Here the bracket $[\cdot,\cdot]$ denotes the commutator of two
matrices.

These relations are complicated.  In practical applications, one
usually looks for higher-order effects only in the QCD coupling, so
that many terms can be dropped.  Also the resummation of logarithms is
then simplified.

\subsection*{Example}

To demonstrate the method, we take the process $H\to b\bar b$ which
has been investigated in ref.~\cite{KS94a}.  Let us consider the
contribution of the diagram in Fig.~\ref{fig1} in the \ms\ scheme,
together with its counterterms (Fig.~\ref{fig2}).  By $A$ ($B,\ldots
E$) we denote the residue of the $1/\epsilon$ pole, where the
space-time dimension is $D=4-\epsilon$.  We consider only the
vectorial part of these self-energy diagrams, which is proportional to
$\slash{p}_b$.  The analysis of the parts proportional to $m_b$, and
of the other diagrams contributing in the given order, proceeds along
similar lines.

The relevant formula is (\ref{zeta-expansion}).  Three coupling
constants are around: $g_b$, $g_t$ (Yukawa couplings), and $g_s$ (QCD
coupling).  However, terms proportional to $g_b$ do not contribute to
the wave-function renormalization in leading order, so that all
expressions in our example are proportional to $g_s^2 g_t^2$.  The
counterterm diagrams correspond to renormalizations of $g_s$ and
$g_t$.  Of course, the net renormalization of $g_s$ is zero here, due
to the QCD Ward identities, but the compensating terms are provided by
different diagrams.

First, we collect the terms which are themselves of order $g_s^2 g_t^2$
and have to be inserted into (\ref{SHT-MS-full}) at tree level.  The
following terms are nonvanishing:
\begin{eqnarray}
  g_s^2 g_t^2 \gamma_2^{ttss}/2 &=& (2A - 2B - 2C)/2,
  \\
  - g_s^2 g_t^2 \hat\gamma_2^{ttss}/2 &=& -(2D)/2,
  \\
  g_s^2 g_t^2 \Delta\zeta_1^{tt}\beta_{t1}^{ss} &=& C/2 \label{deltazeta}.
\end{eqnarray}
For the calculation of $\gamma_2$ we had to subtract the counterterms
for the UV divergent subdiagrams ($B$ and $C$) twice.  The effective
theory term $\hat\gamma_2$ contributes one diagram of order $g_s^2$,
with an insertion of the matching contribution proportional to
$g_t^2$.  The matching term $\Delta\zeta$ is proportional to $g_t^2$
and has to be multiplied with one contribution to the $g_t$ beta
function; this is easily seen to be again equal to $C/2$.  The
prefactor in the left-hand side of (\ref{deltazeta}) is $1$, not $2$,
since $2\Delta\zeta_1\beta_1$ has to be distributed among
Fig.~\ref{fig1} and its mirror image.

Furthermore, we have to consider a contribution of
(\ref{coupling-expansion}) in order $g_t^2$, inserted into the master
equation (\ref{SHT-MS-full}) and evaluated in the effective theory to
order $g_s^2$.  This is equal to
\begin{equation}
  \frac{d\hat g_s}{d\ln m_h}\frac{\partial}{\partial \hat g_s}
	\hat\Gamma^{(2)} = B.
\end{equation}
Taking everything together, the sum of the terms pertaining to
Fig.~\ref{fig1}, inserted into (\ref{SHT-MS-full}) with $n_b=2$, is
\begin{eqnarray}
  2A - B - C - 2D,
\end{eqnarray}
We have assumed that all masses are kept nonzero.

This may be compared with the approach of ref.~\cite{KS94a}.  There
the light masses are neglegted, and the resulting IR divergences are
regulated dimensionally.  This implies that $B$ and $D$ are
identically zero, and their contributions are absorbed into $A$.  The
calculation of Fig.~\ref{fig1}, and the subsequent differentiation
with respect to $m_t^0$ gives a contribution $2A$.  The
renormalization of $g_s$ and $g_t$ (the former being in fact obsolete)
is equivalent to subtracting $B$ and $C$ once.  Thus we have again
the result
\begin{eqnarray}
  2A - B - C - 2D.
\end{eqnarray}
If the on-shell scheme is used, $E$ is also subtracted.

{}From the practical point of view, the latter algorithm looks simpler.
However, the fact that UV and IR divergences are not separated causes
potential problems.  In the first approach the expressions
$\gamma_2-\hat\gamma_2$ and $\Delta\zeta$ are manifestly free of IR
divergences, so that we might as well set $m_b=m_\phi=0$ in their
calculation.  However, the expression $B$ alone is not, so that in the
second approach one has to rely on a QCD Ward identity to ensure that
$B$ is cancelled by other diagrams, and one gets a finite answer.  A
similar statement holds true for the other counterterm diagrams.  If
$C$ is na{\"\i}vely calculated, the corresponding diagram
(Fig.~\ref{fig3}) is UV and IR divergent, and if both divergences are
regulated dimensionally, the relevant coefficient is left
undetermined.  Again, an electroweak Ward identity provides the
correct renormalization of $g_t$, which has been used in
ref.~\cite{KS94a}.

To conclude, the complete effective-theory expressions are somewhat
cumbersome, but they allow a safe diagram-by-diagram analysis.  In
order to simplify the calculations, when additional information is
used (symmetry arguments, explicit expressions for bare parameters),
and IR divergences are controlled, shortcuts are possible.  However,
once logarithms have to be resummed, one has to return to the complete
expressions developed in the main part of the present paper.

\vfill\newpage
\expandafter\ifx\csname fmffile\endcsname\relax\else
\begin{fmffile}{higgs-fig}

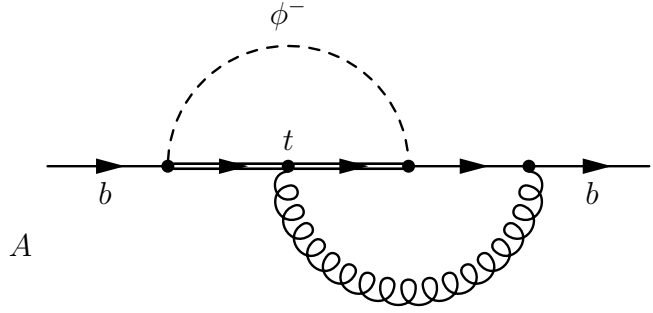
\begin{figure}
\begin{center}
\begin{fmfgraph*}(80,30)
  \fmfleft{i}  \fmfright{o}
  \fmf{fermion,label=$b$,label.side=right}{i,v1}
  \fmf{heavy}{v1,v2}
  \fmfv{label=$t$,label.angle=90}{v2}
  \fmf{heavy}{v2,v3}
  \fmf{fermion}{v3,v4}
  \fmf{fermion,label=$b$,label.side=right}{v4,o}
  \fmf{dashes,label=$\phi^-$,left,tension=0}{v1,v3}
  \fmf{gluon,right,tension=0}{v2,v4}
  \fmfdotn{v}{4}
  \fmfforce{(0,5mm)}{a} \fmfv{label=$A$}{a}
\end{fmfgraph*}
\end{center}
\caption{Sample diagram in a NLO calculation: The coefficient of the
$1/\epsilon$ pole is denoted by~$A$.}
\label{fig1}
\end{figure}

\begin{figure}
\begin{center}
\begin{fmfgraph*}(40,30)
  \fmfleft{i}  \fmfright{o}
  \fmf{fermion}{i,v1}
  \fmf{fermion}{v1,v2}
  \fmf{fermion}{v2,o}
  \fmf{gluon,right,tension=0}{v1,v2}
  \fmfv{decor.shape=square,decor.filled=0,decor.size=4mm}{v1}
  \fmfdot{v2}
  \fmfforce{(0,10mm)}{a}  \fmfv{label=$B$}{a}
\end{fmfgraph*}
\hskip2cm
\begin{fmfgraph*}(40,30)
  \fmfleft{i}  \fmfright{o}
  \fmf{fermion}{i,v1}
  \fmf{heavy}{v1,v2}
  \fmf{fermion}{v2,o}
  \fmf{dashes,left,tension=0}{v1,v2}
  \fmfv{decor.shape=square,decor.filled=0,decor.size=4mm}{v2}
  \fmfdot{v1}
  \fmfforce{(0,10mm)}{a}
  \fmfv{label=$C$}{a}
\end{fmfgraph*}
\\
\begin{fmfgraph*}(40,30)
  \fmfleft{i}  \fmfright{o}
  \fmf{fermion}{i,v1}
  \fmf{fermion}{v1,v2}
  \fmf{fermion}{v2,o}
  \fmf{gluon,right,tension=0}{v1,v2}
  \fmfv{decor.shape=square,decor.filled=1,decor.size=4mm}{v1}
  \fmfdot{v2}
  \fmfforce{(0,10mm)}{a}  \fmfv{label=$D$}{a}
\end{fmfgraph*}
\hskip2cm
\begin{fmfgraph*}(40,30)
  \fmfleft{i}  \fmfright{o}
  \fmf{fermion}{i,v1}
  \fmf{heavy}{v1,v2}
  \fmf{fermion}{v2,o}
  \fmf{dashes,left,tension=0}{v1,v2}
  \fmfv{decor.shape=square,decor.filled=1,decor.size=4mm}{v2}
  \fmfdot{v1}
  \fmfforce{(0,10mm)}{a}  \fmfv{label=$E$}{a}
\end{fmfgraph*}
\end{center}
\caption{Counterterm diagrams: $B$, $C$, $D$, and $E$ are the
coefficients of the $1/\epsilon$ poles.  An open square denotes the
pole part of the subdiagram; a filled square stands for the finite part.}
\label{fig2}
\end{figure}
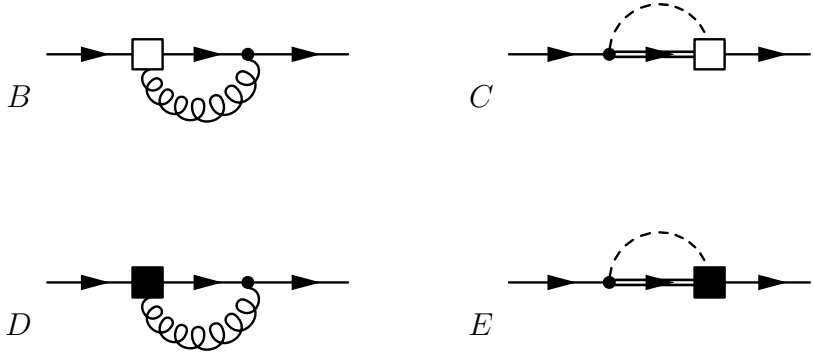

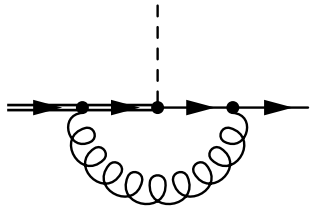
\begin{figure}
\begin{center}
\begin{fmfgraph}(40,30)
  \fmfsurround{o,v,i,x}
  \fmf{heavy}{i,v1}
  \fmf{heavy}{v1,v2}
  \fmf{fermion}{v2,v3}
  \fmf{fermion}{v3,o}
  \fmf{dashes,tension=0}{v2,v}
  \fmf{gluon,right,tension=0}{v1,v3}
  \fmfdotn{v}{3}
\end{fmfgraph}
\end{center}
\caption{Subdiagram for $C$ and $E$.}
\label{fig3}
\end{figure}

\end{fmffile}
\fi

\end{document}